\documentclass[twoside]{article}
\usepackage{epsfig.sty}
\newcommand{\beq}{\begin{eqnarray}}
\newcommand{\eeq}{\end{eqnarray}}
\begin{document}
\title{Top Quark Mixed Hybrid Meson and the Quark-Gluon Plasma}
\author{Leonard S. Kisslinger\\
Department of Physics, Carnegie Mellon University, Pittsburgh PA 15213 USA.\\
Debasish Das\\
Saha Institute of Nuclear Physics,HBNI, Bidhan Nagar, Kolkata 700064, INDIA.}
\date{}
\maketitle 
\begin{abstract} Using the method of QCD Sum Rules we estimate the energy of
lowest energy top quark meson state with a hybrid admixture. This new
estimate uses the previous estimates of hybrid charmonium and upsilon states, 
but with the  mass of the top quark mass being much greater than the charm and 
bottom quark masses. We discuss production of mixed hybrid top quark mesons and
possible detection of the creation of the Quark-Gluon Plasma via Relativistic
Heavy Ion Collisions.

\end{abstract}
\maketitle
\noindent
PACS Indices:12.38.Aw,13.60.Le,14.40.Lb,14.40Nd
\vspace{1mm}


\section{Introduction}
  
  In previous estimates of mixed heavy quark hybrid mesons\cite{kpr08,lsk09}
the charmonium meson state $\Psi(2S)$ and upsilon meson state $\Upsilon(3S)$
state were shown to be approximately 50\% standard meson and 50\% hybrid meson,
with the hybrid meson having an active gluon. QCD Sules were also used to
estimate the mass of a scalar glueball\cite{kj01}.

The quarks associated with
$\Psi(2S),\Upsilon(3S)$ are the charm and bottom quarks with masses $M_c 
\simeq 1.27$ GeV, $M_b \simeq 4.18$ GeV. Since the top quark mass, 
$M_t\simeq 173$ GeV\cite{ppb16} is much larger than the charm, bottom quark 
masses, $M_c,M_b$, top meson states have not been
detected. Recently, however a signal for $t\bar{t}$ production in p-Pb 
collisions with $\sqrt{s_{NN}}=$ 8.16 TeV has been observed\cite{CMS17}.
Also, an important motivation for the present work is the recent study of
the possible determination of the creation of the Quark-Gluon Plasma (QGP)
via the production of $t\bar{t}$ events\cite{amsS18}.

In order to estimate the energy of lowest energy hybrid top quark meson state 
we need the quark and gluon condensates. Note that FIG. 8 in Ref.\cite{lsk09}
gave the dominant diagram so the charm quark condenate $<c\bar{c}>$ and bottom
quark condensate $<b\bar{b}>$ were not included in the estimates of mixed
charmonium and bottomonium hybrid mesons.

However, the top quark condensate, $<t\bar{t}>$, must be included in the
present work. From Ref.\cite{panpz13} $ <t\bar{t}> \simeq
(126 {\rm GeV})^3$, which is used to estimate the mixed top quark meson
correlator in section 3. Another important parameter to estimate the
mass of a  mixed top quark hybrid meson is the gluon condensate $<G^2>$,
whose value is discussed and given below.

In the next section we give a brief review of the  method of QCD Sum Rules,
and in the following section we estimate the energy of lowest energy mixed
hybrid top quark meson state using QCD Sum Rules. As with the charm and bottom 
$\Psi(nS),\Upsilon(nS)$states, we choose a top current so the $\Xi(nS)$ states
are $1^{--}$ states. Note that in Refs\cite{kpr08,lsk09} the $\Psi(2S),
\Upsilon(3S)$ were found to be mixed hybrid states, while in the present
research we find that the $\Xi(1S)$ is a mixed hybrid top meson state.

Since the top quark condensate, $<t\bar{t}>$ is large we must also
include the top quark condensate correlator, $\Pi_{Hc}^{\mu \nu }(p)$, which was
not necessary in estimates of the mixed hybrid $\Psi(nS),\Upsilon(nS)$states.

Also, after our derivation of the correlator of the mixed state,  
$\Pi_{H-HH}^{\mu\nu}(p)$, we find there are terms needing Borel transforms that
were not needed in Refs\cite{kpr08,lsk09}.

\newpage

\section{Brief Review of theMethod of QCD Sum Rules}
 The starting point of the method of QCD sum rules\cite{sz79} for finding 
$M_A$, the mass of a state A, is the correlator,
 
\beq
\label{2}
       \Pi^A(x) &=&  \langle | T[J_A(x) J_A(0)]|\rangle \; ,
\eeq
with $| \rangle$ the vacuum state and
the current $J_A(x)$ creating the states with quantum numbers A:
\beq
\label{3}
     J_A(x)| \rangle &=& c_A |A \rangle + \sum_n c_n |n; A  \rangle  \; ,
\eeq
where $ |A \rangle$ is the lowest energy state with quantum numbers A.
After a Fourier transform to momentum space,
\beq 
\label{4}
       \Pi^A(x) &\Rightarrow&\Pi^A(p) \; ,
\eeq
where $\vec{p}$ is the momentum. The Borel Transform\cite{sz79}, is defined by

\beq
\label{B}
 \mathcal{B}_{M_B^2}&=& lim_{p^2,n \rightarrow \infty;\frac{p^2}{n}=M_B^2}
\frac{(p^2)^{n+1}}{n!}(\frac{-d}{dp^2})^n \; ,
\eeq
with
\beq
\label{5}
     \mathcal{B}_{M_B^2}\Pi^A(p)&=&  \Pi^A(M_B) \; ,
\eeq
with $M_B$ the Borel mass. Defing the right-hand side (rhs) of the sum rule
\beq
\label{6}
  \Pi(p)_{\rm{rhs}}^A &=& \sum_k c_k(p) \langle 0|{\cal O}_k|0\rangle
 \; ,
\eeq
where $c_k(p)$ are the Wilson coefficients and $\langle 0|{\cal O}_k|0\rangle$
are gauge invariant operators constructed from quark and gluon fields,
the final QCD sum rule is
\beq
\label{7}
  \Pi^A(M_B)&=& \mathcal{B}_{M_B^2} \Pi(p)_{\rm{rhs}}^A \; .
\eeq

  The value of $M_A$ is found as the minimum in the plot of $M_A$ vs $M_B$.

\section{Mixed Top Quark $1^{--}$ States Using QCD Sum Rules}

 Based on previous work\cite{lsk09} we assume that there is strong mixing 
between a top quark meson and a hybrid top quark meson with the same quantum 
numbers (as shown below).

\subsection{Top Quark Meson Correlator}

 We now attempt to find the lowest $J^{PC}=1^{--}$ 
state with a sizable admixture of a top meson and a hybrid top meson.
The mixed vector ($J^{PC}=1^{--}$) top, hybrid top current we use in 
QCD Sum Rules is
\beq
\label{11}
        J^\mu &=& b J_H^\mu + \sqrt{1-b^2} J_{HH}^\mu 
\eeq
with
\beq
\label{12}
          J_H^\mu &=&  \bar{\Psi_t} \gamma^\mu \Psi_t \nonumber \\
          J_{HH}^\mu&=&  \bar{\Psi_t}\Gamma_\nu G^{\mu\nu} \Psi_t \; ,
\eeq
where $\Psi_t$ is the top quark field, $\Gamma_\nu = C \gamma_\nu$,
$\gamma_\nu$ is the usual Dirac matrix, C is the charge conjugation operator,
and the gluon color field is
\beq
\label{13}
         G^{\mu\nu}&=& \sum_{a=1}^8 \frac{\lambda_a}{2} G_a^{\mu\nu} \; ,
\eeq
with $\lambda_a$ the SU(3) generator ($Tr[\lambda_a \lambda_b]= 2 \delta_{ab}$).
\newpage
Therefore the correlator for the mixed state:
\beq
\label{14} 
   \Pi_{H-HH}^{\mu\nu}(x) &=& <0|T[J^\mu(x) J^\nu(0)]|0>
\eeq
is 
\beq
\label{15}
   \Pi_{H-HH}^{\mu\nu}(x) &=& b^2  \Pi_{H}^{\mu\nu}(x) + (1-b^2)
\Pi_{HH}^{\mu\nu}(x) \nonumber \\
  && +2b\sqrt{1-b^2}\Pi_{HHH}^{\mu\nu}(x) \\
       \Pi_{H}^{\mu\nu}(x)&=& <0|T[J_H^\mu(x) J_H^\nu(0)]|0> \nonumber \\
   \Pi_{HH}^{\mu\nu}(x)&=& <0|T[J_{HH}^\mu(x) J_{HH}^\nu(0)]|0> \nonumber \\
   \Pi_{HHH}^{\mu\nu}(x)&=& <0|T[J_H^\mu(x) J_{HH}^\nu(0)]|0> \nonumber \; ,
\eeq
where $\Pi_{H}^{\mu\nu}, \Pi_{HH}^{\mu\nu}, Pi_{HHH}^{\mu\nu}$ are the correlators for
the top quark meson, the hybrid top quark meson, the mixed meson and hybrid
meson.

  To stimate the mass of the mixed meson and hybrid meson we need the top
quark condensate and the gluon condensate. The estimated value of the top 
quark condensate is\cite{panpz13}
\beq
\label{topcond}
 <t\bar{t}> &\simeq& (126 {\rm GeV})^3 \; .
\eeq

 In Refs\cite{tf95,lz95} $<\alpha_s G^2/\pi>$ were estimated, where $\alpha_s$ 
is the strong coupling constant and $\alpha_s/\pi \simeq 0.57$. Using this
value of $\alpha_s/\pi$, from  Refs\cite{tf95,lz95} the gluon condensate is
\beq
\label{gluecond}
  <G^2> &\simeq& 0.003 GeV^4 \; .
\eeq

  The basic diagrams for the top quark meson correlator are shown in Figure 1

\begin{figure}[ht]
\begin{center}
\epsfig{file=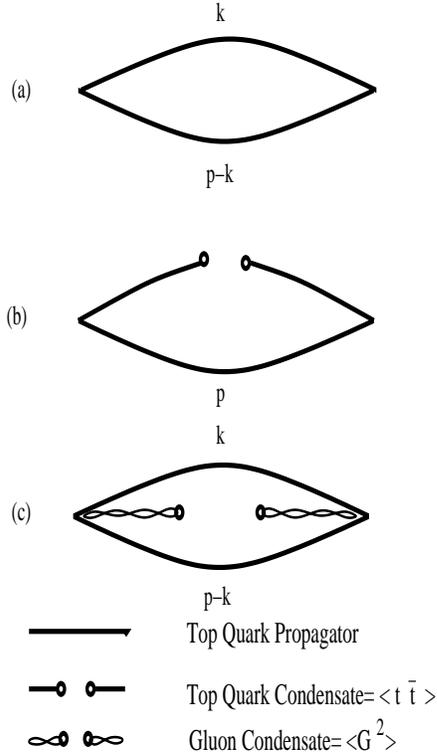,height=10cm,width=6cm}
\caption{Heavy quark meson diagrams: 
(a) standard correlator;  (b) Quark Condensate; (c) Gluon Condensate}
\label{Fig.1}
\end{center}
\end{figure}
\newpage

  From Eqs(\ref{12},\ref{15}) the quark correlator in momentum space,
corresponding to Figure 1 (a), is
\beq
\label{Hcorrelator}
  \Pi_{H}^{\mu \nu }(p)& =& g_v^2\int \frac{d^4 k}{(2 \pi)^4} 
Tr[S(k) \Gamma_5^\mu S(p-k)\Gamma_5^{\nu T}] \nonumber \\
          S(k) &=& \frac{\not\!k + M_t}{k^2-M_t^2} \\
        \Gamma_5^\mu &=& \gamma^\mu \gamma_5 \nonumber \\
Tr[S(k) \Gamma_5^\mu S(p-k)\Gamma_5^{\nu T}]&=& \frac{M_t^2 Tr[\gamma^\mu
\gamma^\nu]-k_\alpha (p-k)_\beta Tr[\gamma^\alpha \gamma^\mu \gamma^\beta 
\gamma^\nu]}{(k^2-M_t^2)[(p-k)^2-M_t^2]} \; .
\eeq

The integrals needed to evaluate Eq(\ref{Hcorrelator}) are
\beq
\label{k-integrals}
 \int \frac{d^4 k}{(2 \pi)^4}\frac{1}{(k^2-M_t^2)[(p-k)^2-M_t^2]}
&=& \frac{2M_t^2-p^2/2}{(4 \pi)^2}  I_o(p) \\
\int \frac{d^4 k}{(2 \pi)^4}\frac{k^\mu}{(k^2-M_t^2)[(p-k)^2-M_t^2]}
&=& \frac{p^\mu}{(4 \pi)^2}[(M_t^2-p^2/4)I_o(p) +7/4] \nonumber \\
\int \frac{d^4 k}{(2 \pi)^4}\frac{k^\mu k^\nu}{(k^2-M_t^2)[(p-k)^2-M_t^2]}
&=& \frac{g^{\mu \nu} p^2}{12(4 \pi)^2}(M_t^2-\frac{p^2}{2}) I_o(p)+
\frac{p^\mu p^\nu}{(4 \pi)^2}[(\frac{5M_t^2-p^2}{6}-\frac{2M_t^4}{p^2})I_o(p)-
\frac{2M_t^2}{3p^2}] \nonumber
\eeq
with 
\beq
\label{IoP}
  I_o(p)&=&\int_{0}^{1}\frac{d \alpha}{\alpha(1-\alpha)p^2-M_t^2} \; .
\eeq

  Carrying out the traces and the $k$ integral and dropping terms that vanish
with a Borel transform

\beq
\label{Hcor-final}
  \Pi_{H}^{\mu \nu }(p)& =& g^{\mu \nu } \frac{4 g_v^2}{(4 \pi)^2} 
p^2(M_t^2-\frac{p^2}{4}) I_o(p) \; . 
\eeq

  With the scalar correlator $\Pi^S$  defined by $\Pi^{\mu \nu}(p)=
(p_\mu p_\nu/p^2-g^{\mu \nu} )\Pi^V(p)+ (p_\mu p_\nu/p^2) \Pi^S(p)$, the
scalar correlator $\Pi_{H}^{S}(p)$ is
\beq
\label{H-scalar}
  \Pi_{H}^{S}(p)& =& \frac{4 g_v^2}{(4 \pi)^2}p^2(M_t^2-\frac{p^2}{4}) I_o(p)
\; .
\eeq

 The top quark condensate correlator corresponding to Figure 1 (b), 
 $\Pi_{Hc}^{\mu \nu }(p)$, with the condensate carrying no momentum so
$ S_c(k) =-<t\bar{t}> M_t \delta(k-0){-M_t^2}$, and
\beq
\label{H-cond}
\Pi_{Hc}^{\mu \nu }(p)&=&-\frac{g_v^2}{(2 \pi)^4}<t\bar{t}> M_t
4g^{\mu \nu}\frac{M_t}{p^2-M_t^2} \nonumber \\
 &=& -4g^{\mu \nu}\frac{g_v^2}{(2 \pi)^4}\frac{<t\bar{t}> M_t}{p^2-M_t^2} \; .
\eeq
 
Therefore the scalar top quark condensate correlator is
\beq
\label{H-cond-scalar}
\Pi_{Hc}^{S}(p)&=&-4\frac{g_v^2}{(2 \pi)^4}\frac{<t\bar{t}> M_t}{p^2-M_t^2} \; .
\eeq

 Noting from Eq(\ref{gluecond}), as in Ref\cite{lsk09}, we do not include 
the Gluon Condensate term, Figure 1 (c).

\subsection{Top Quark Hybrid Meson Correlator}

\begin{figure}[ht]
\begin{center}
\epsfig{file=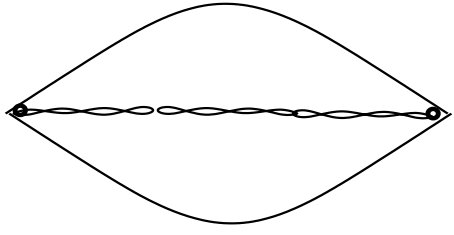,height=3cm,width=6cm}
\caption{Top Quark Hybrid Correlator}
\label{Fig.5}
\end{center}
\end{figure}

  From Eqs(\ref{12},\ref{13}) the top quark hybrid correlator\cite{kpr08,lsk09}
\beq
\label{HHcorrelator}
  \Pi_{HH}^{\mu \nu }(p)& =&6ig_v^2\int \frac{d^4 k}{(2 \pi)^4} 
Tr[S(k) \Gamma_5^\alpha S(p-k)\Gamma_5^{\beta T}]Tr[G^{\mu \alpha} 
G^{\nu \beta}](k) \; ,
\eeq
with
\beq
\label{G-G}
 Tr[G^{\mu \alpha} G^{\nu \beta}](k)&=&-i4\pi^2(g^{\alpha \beta}
\frac{k^\mu k^\nu}{k^2} +g^{\mu \nu}\frac{k^\alpha k^\beta}{k^2}
-g^{\mu \beta}\frac{k^\alpha k^\nu}{k^2} -
g^{\alpha \nu}\frac{k^\mu k^\beta}{k^2}) \; .
\eeq

In order to carry out the integrals for $\Pi_{HH}^{\mu \nu }(p)$, in addition to
the integrals in Eq(\ref{k-integrals}) we need
\beq
\label{k-integrals-2}
 \int \frac{d^4 k}{(2 \pi)^4}\frac{k^2}{(k^2-M_t^2)[(p-k)^2-M_t^2]}&=&
 M_t^2\frac{2M_t^2-p^2/2}{(4 \pi)^2} I_o(p)+\rm{ind\;of\;p}
\eeq

 \beq
\label{k-integrals-3}\int \frac{d^4 k}{(2 \pi)^4}\frac{k^\mu k^\nu}
{k^2(k^2-M_t^2)[(p-k)^2-M_t^2]}
&=&\frac{1}{(4 \pi)^2}\int_{0}^{\infty} d\rho \frac{\delta(1-\lambda)}{\rho}
\int_{0}^{1}d\alpha\int_{0}^{1}d\beta\int_{0}^{1}d\gamma(\frac{g^{\mu \nu}}{8}
+\frac{p^\mu p^\nu \rho \gamma^2}{\lambda})e^{-p^2 \rho(\gamma-\gamma^2)}
\nonumber
\eeq

\beq
\label{k-integrals-4}
\int\frac{d^4 k}{(2\pi)^4}\frac{k^\mu k^\nu k\cdot p}{k^2(k^2-M_t^2)[(p-k)^2-
M_t^2]} = 
 -\frac{1}{2(4 \pi)^2}[\frac{g^{\mu \nu}p^2}{6}
(M_t^2-\frac{p^2}{2}) I_o(p)+p^\mu p^\nu[(\frac{5M_t^2-p^2}{6}-\frac{2M_t^4}
{p^2})I_o(p)-\frac{2M_t^2}{3p^2}] \nonumber \\
+ \frac{M_t^2-p^2}{2(4 \pi)^2}\int_{0}^{\infty} d\rho 
\frac{\delta(1-\lambda)}{\rho}
 \int_{0}^{1}d\alpha\int_{0}^{1}d\beta
\int_{0}^{1}d\gamma(\frac{g^{\mu \nu}}{8} +\frac{p^\mu p^\nu \rho \gamma^2}
{\lambda})e^{-p^2 \rho(\gamma-\gamma^2)} \nonumber
\eeq

From Eqs(\ref{HHcorrelator},\ref{G-G},\ref{k-integrals},\ref{k-integrals-2})
\beq
\label{HHcorrelator-2}
 \Pi_{HH}^{\mu \nu }(p)= 6 g_v^2[g^{\mu \nu}I_o(p)(16M_t^4-39M_t^2 p^2+ p^4/2)
+p^\mu p^\nu I_o(p)(\frac{63 M_t^2}{6}-\frac{19 p^2}{2}-10\frac{M_T^4}{p^2})
 -p^\mu p^\nu (\frac{28 M_t^2}{3 p^2} +7) \nonumber \\
-(24M_t^2-8p^2)\int_{0}^{\infty} d\rho \frac{\delta(1-\lambda)}{\rho}
\int_{0}^{1}d\alpha\int_{0}^{1}d\beta\int_{0}^{1}d\gamma(\frac{g^{\mu \nu}}{8}
+\frac{p^\mu p^\nu \rho \gamma^2}{\lambda})e^{-p^2 \rho(\gamma-\gamma^2)}]
\eeq

 Therefore, using $\Pi^{\mu \nu}(p)= (p_\mu p_\nu/p^2-g^{\mu \nu} )\Pi^V(p)+ 
(p_\mu p_\nu/p^2) \Pi^S(p)$ 
\beq
\label{HHcorrelator-scalar}
 \Pi_{HH}^{S}(p)=-i g_v^2[(6M_t^4-28.5M_t^2p^2-9p^4)I_o(p) \\
-(24M_t^2-8p^2)\int_{0}^{\infty} d\rho \frac{\delta(1-\lambda)}{\rho}
(\frac{1}{8}+\frac{p^2 \rho \gamma^2}{\lambda})\int_{0}^{1}d\alpha
\int_{0}^{1}d\beta\int_{0}^{1}d\gamma e^{-p^2 \rho(\gamma-\gamma^2)}]
\nonumber
\eeq

Note that from Ref\cite{kpr08} the Borel transform, $\mathcal{B}$, of
the $I_o(p)$ terms can be found, but the  Borel transform of the terms
$p^{2n} e^{-p^2 \rho(\gamma-\gamma^2)}$ (with n=0,1,2) must be derived.
\newpage

The mixed top quark-top quark hybrid correlator,  $\Pi_{HHH}^{\mu \nu }(p)$,
is shown in figure 3 below. 

\vspace{1cm}
\begin{figure}[ht]
\begin{center}
\epsfig{file=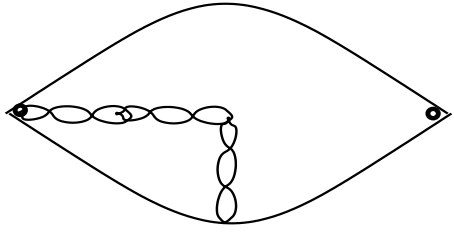,height=3cm,width=6cm}
\caption{Mixed Top Quark, Hybrid Top Quark, Correlator}
\label{Fig.6}
\end{center}
\end{figure}

From Ref\cite{lsk09} 
\beq
\label{HHH-H-scalar}
 \Pi_{HHH}^{S}(p) &\simeq& \pi^2 \Pi_{H}^{S}(p) \; .
\eeq

Therefore from Eqs(\ref{11},\ref{H-scalar},\ref{H-cond-scalar},
\ref{HHcorrelator-scalar},\ref{HHH-H-scalar}), with $b=-0.7$\cite{lsk09}, we 
obtain the scalar correlator $\Pi_{H-HH}^{S}(p)$ 
needed to obtain the mass of a mixed hybrid top quark meson
\beq
\label{final-correlator}
 \Pi_{H-HH}^{S}(p) &=& (1+\pi^2)\Pi_{H}^{S}(p)+\Pi_{Hc}^{S}(p)+ \Pi_{HH}^{S}(p)
 \; .
\eeq

Next we must derive the Borel transform $\mathcal{B} \Pi_{H-HH}^{S}(p)$ to
obtain $\Pi_{H-HH}^{S}(M_B))$ from which we will estimate the mass of a
top quark mixed hybrid meson.

\section{Borel Transform of $\Pi_{H-HH}^{S}(p)$ and Estimate of
the Mass of a  Mixed Hybrid Top Quark Meson}

With the Borel transform defined in Eq(\ref{B})
the Borel transforms needed in this work are:
\beq
\label{BI}
 \mathcal{B}_{M_B^2} I_o(p)&=&2e^{-\frac{2M_t^2}{M_B^2}} 
K_0(\frac{2M_t^2}{M_B^2}) \nonumber\\
 \mathcal{B}_{M_B^2} p^2I_o(p)&=&4M_t^2e^{-\frac{2M_t^2}{M_B^2}}
[ K_0(\frac{2M_t^2}{M_B^2})+
K_1(\frac{2M_t^2}{M_B^2})]\nonumber\\
 \mathcal{B}_{M_B^2} p^4I_o(p)&=&4M_t^4e^{-\frac{2M_t^2}{M_B^2}}
[3 K_0(\frac{2M_t^2}{M_B^2})+
4 K_1(\frac{2M_t^2}{M_B^2})+K_2(\frac{2M_t^2}{M_B^2})]\nonumber\\
 \mathcal{B}_{M_B^2} \frac{1}{p^2-M_t^2}&=&-e^{-\frac{M_t^2}{M_B^2}}\nonumber\\
 \mathcal{B}_{M_B^2}e^{-p^2 \rho(\gamma-\gamma^2)}&=& 0  \; .
\eeq

Therefore from Eqs((\ref{H-scalar},\ref{H-cond-scalar},
\ref{HHcorrelator-scalar},\ref{final-correlator},\ref{B}) $B\Pi_{H-HH}^{S}(p)= 
\Pi_{H-HH}^{S}(M_B)$, with $x=\frac{2M_t^2}{M_B^2}$, is\cite{kpr08}
\beq
\label{final-correlator-Borel}
\Pi_{H-HH}^{S}(x)&\simeq& g_v^2 M_t^4 e^{-x}
[-1,260K_0(-x)-1,548 K_1(-x)-210 K_2((-x)]\nonumber\\
  && + g_v^2 0.0026 <t\bar{t}> M_t e^{-\frac{M_t^2}{M_B^2}} \; .
\eeq
\newpage

  The Modified Bessel functions of the second kind, $K_n(x)$ in 
Eq(\ref{final-correlator-Borel}), are\cite{b58}
\beq
\label{Kn}
K_0(x)&=&\sqrt{\frac{\pi}{2x}}e^{-x}(1-\frac{1}{8x}+\frac{9}{128x^2}
-\frac{225}{3072x^3}) \nonumber \\
K_1(x)&=&\sqrt{\frac{\pi}{2x}}e^{-x}(1+\frac{3}{8x}-\frac{15}{128x^2}
+\frac{315}{3072x^3}) \nonumber \\
K_2(x)&=&\sqrt{\frac{\pi}{2x}}e^{-x}(1+\frac{15}{8x}+\frac{105}{128x^2}
-\frac{945}{3072x^3}) \; .
\eeq

Using $g_v^2\simeq 1.49$\cite{ppb16} and $M_t\simeq$173 GeV, $C\equiv g_v^2
\sqrt{\pi/2}M_t^4 \simeq 1.67\times 10^9 (\rm{ GeV})^4$. 
$D\equiv  g_v^2 0.0026 <t\bar{t}> M_t\simeq 2.22\times 10^5 (\rm{ GeV})^4$.
Since $D \ll C$, we drop the last term in Eq(\ref{final-correlator-Borel})
From Eqs(\ref{final-correlator-Borel},\ref{Kn}), using $\sqrt{-x}=i\sqrt{x}$,
\beq
\label{H-HH-correlator-Borel}
\Pi_{H-HH}^{S}(x)&\simeq& -\frac{C}{i}\sqrt{\frac{1}{x}} e^{x}
(3018-\frac{816.75}{x}+\frac{79.45}{x^2}-\frac{1.85}{x^3}) \; .
\eeq

As in Refs\cite{kpr08,lsk09}, The mass of the top quark mixed hybrid meson,
$M_{tHH}$ is obtained from the ratio of the derivitave of $\Pi_{H-HH}^{S}(M_B)$
with respect to $1/M_{B^2}$ to $\Pi_{H-HH}^{S}(M_B)$:
\beq
\label{derative}
  M_{tHH}^2&=& \frac{d}{d(1/M_B^2)}\Pi_{H-HH}^{S}(M_B)/\Pi_{H-HH}^{S}(M_B) \; .
\eeq

Since $x=\frac{2M_t^2}{M_B^2}$, or $\frac{d}{d(1/M_B^2)}=2M_t^2\frac{d}{dx}$
\beq
\label{MtHH}
  M_{tHH}^2 &=& 2M_t^2 \frac{d}{dx}\Pi_{H-HH}^{S}(x)/\Pi_{H-HH}^{S}(x) \; .
\eeq

From Eqs(\ref{MtHH},\ref{H-HH-correlator-Borel})
\beq
\label{M2tHH}
  M_{tHH}^2 &=& 2M_t^2\frac{3018-2325.8\frac{1}{x}+1304.6\frac{1}{x^2}
-200.6\frac{1}{x^3}+6.48\frac{1}{x^4}}{3018-816.75\frac{1}{x}+79.45\frac{1}{x^2}
-1.85\frac{1}{x^3}} \; .
\eeq

In Figure 4 below $M_{tHH}^2$ is plotted vs $M_B^2=2M_t^2/x$
\vspace{1cm}
\begin{figure}[ht]
\begin{center}
\epsfig{file=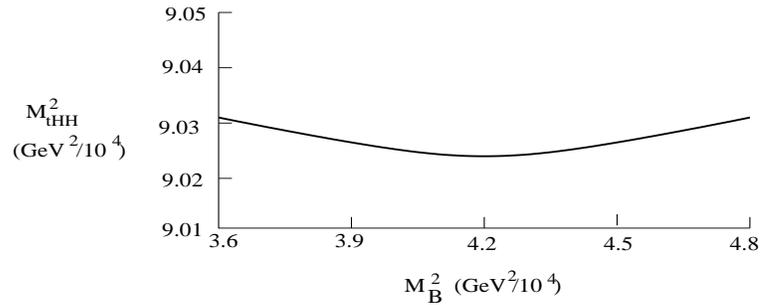,height=4cm,width=10cm}
\caption{Mixed top-hybrid top mass approximately 300 GeV}
\label{Fig.4}
\end{center}
\end{figure}
\newpage
\section{Results and Conclusions}

As discussed in Ref\cite{lsk17} although the detection of the Quark-Gluon 
Plasma (QGP) produced in Relativistic Heavy Ion Collisions (RHIC) is difficult, 
the production of mixed heavy quark hybrid mesons via RHIC is a possible
mechanism for detecting the QGP. In Ref\cite{lsk17} the mixed heavy quark 
hybrid charmonium meson state $\Psi(2S)$ and upsilon meson state $\Upsilon(3S)$
were considered. As discussed above  $t\bar{t}$ production in p-Pb  has been 
observed\cite{CMS17}, and there as a possible determination of the creation of 
the QGP via the production of $t\bar{t}$ events\cite{amsS18}. The main result
of our present work it that there is a mixed top quark hybrid meson state with
a mass approximately 300 GeV, and with the increased energy at the LHC the 
production of this state could be used to detect the QGP. 
 
\vspace{5mm}
\Large
{\bf Acknowledgements}
\normalsize
\vspace{5mm}

Author D.D. acknowledges the facilities of Saha Institute of Nuclear Physics, 
Kolkata, India. Author L.S.K. acknowledges support in part by a grant from
the Pittsburgh Foundation.

\end{document}